\documentclass{icrc29}
\usepackage{graphicx,amssymb,amsmath,times}
\begin{document}
\title[Search for a diffuse flux...]  
{Search for a diffuse flux of high-energy 
neutrinos with the NT200 neutrino telescope } 

\author[Aynutdinov et al ...]  { 
    V. Aynutdinov$^a$, V.Balkanov$^a$, I. Belolaptikov$^g$, N.Budnev$^b$,
    L. Bezrukov$^a$, D. Borschev$^a$,
\newauthor 
    A.Chensky$^b$, I. Danilchenko$^a$, 
    Ya.Davidov$^a$,Zh.-A. Djilkibaev$^a$, G. Domogatsky$^a$, 
\newauthor
A.Dyachok$^b$, 
    S.Fialkovsky$^d$,O.Gaponenko$^a$, O. Gress$^b$, T. Gress$^b$, O.Grishin$^b$, 
    A.Klabukov$^a$, 
\newauthor 
    A.Klimov$^f$, K.Konischev$^g$, A.Koshechkin$^a$, 
    L.Kuzmichev$^c$, V.Kulepov$^d$, 
\newauthor  
    B.Lubsandorzhiev$^a$, S.Mikheyev$^a$, 
    M.Milenin$^d$, R.Mirgazov$^b$, T.Mikolajski$^h$, E.Osipova$^c$,
\newauthor 
    A.Pavlov$^b$,  G.Pan'kov$^b$, L.Pan'kov$^b$, A.Panfilov$^a$,
    \framebox{Yu.Parfenov$^b$}~, D.Petukhov$^a$,
\newauthor 
    E.Pliskovsky$^g$, P.Pokhil$^a$, V.Polecshuk$^a$, E.Popova$^c$, V.Prosin$^c$, 
    M.Rozanov$^e$, 
\newauthor 
    V.Rubtzov$^b$, B.Shaibonov$^a$, A.Shirokov$^c$, Ch.Spiering$^h$, 
    B.Tarashansky$^b$,
 \newauthor 
    R.Vasiliev$^g$, E.Vyatchin$^a$, R.Wischnewski$^h$,
    I.Yashin$^c$, V.Zhukov$^a$ \\ 
 (a) Institute for Nuclear Reseach, Russia\\
 (b) Irkutsk State University,Russia\\
 (c) Skobeltsin Institute of Nuclear Physics, Moscow State University, Russia\\
 (d) Nizni Novgorod State Technical University, Russia\\
 (e) St.Petersburg State Marine Technical University, Russia\\
 (f) Kurchatov Institute, Russia\\
 (g) Joint Institute for Nuclear Research, Dubna, Russia\\
 (h) DESY, Zeuthen, Germany}

\presenter{Presenter: R. Wischnewski (ralf.wischnewski@desy.de), \  
ger-wischnewski-R-abs1-og25-oral}

\maketitle

\begin{abstract}

We present the results of a search for high energy extraterrestrial neutrinos
with the Baikal underwater Cherenkov detector {\it NT200}, 
based on data taken in 1998 - 2002 (1038 live days).
Upper limits on the diffuse fluxes of
$\nu_e+\nu_{\mu}+\nu_{\tau}$, predicted by several models 
of AGN-like neutrino sources, are derived.  
For  an $E^{-2}$ behavior of the neutrino spectrum,
our limit is
 $E^2 \Phi_{\nu}(E)<8.1\times 10^{-7}\,
\mbox{cm}^{-2}\,\mbox{s}^{-1}\,
\mbox{sr}^{-1}\,\mbox{GeV}$ 
over an neutrino energy range $2\times10^4 \div 5 \times 10^7\,\mbox{GeV}$
covering 90\% of expected events.
The upper limit on the resonant $\bar{\nu}_e$ diffuse flux is 
$\Phi_{\bar{\nu}_e}<$3.3$\times$10$^{-20}$ 
cm$^{-2}$s$^{-1}$sr$^{-1}$GeV$^{-1}$.

\end{abstract}

\section{Introduction}

High energy neutrinos are likely produced in many violent processes 
in the Universe. Their detection would unambiguously reveal the hadronic
nature of the underlying processes. Neutrinos would be generated by 
proton-proton or proton-photon interactions followed by  production 
and decay of charged mesons. 

A description of the Baikal-detector as well
as physics results from data collected in the years 1998 - 2000
have been presented elsewhere \cite{APP1,APP2,NU04,IC05_stat,IC05_mon}. 
In this paper we present new results of a search for diffuse neutrinos with energies 
larger than 10 TeV.  
The analysis is based on data taken with the Baikal 
neutrino telescope {\it NT200} in the years 1998-2002. Instead of focusing 
to particles crossing the array,  the  analysis is tailored 
to signatures of isolated high-energy cascades in a large volume around the 
detector. This search strategy dramatically enhances the sensitivity of 
{\it NT200} to diffuse high energy processes.

The cascades can stem from leptons and hadrons produced in high energy 
charged current processes.
Obviously, the energy released by the hadronic 
cascade in NC-reactions is small compared to that of the leptonic cascades 
in CC-reactions. Since only electrons and taus develop cascades (the one by directly 
showering up, the other via its decay to secondary particles which develop 
a cascade), the sensitivity of this search is dominated by  $\nu_e$ and 
$\nu_{\tau}$ detection.

\section{Data selection and analysis}

Within the 1038 days of the detector live time
between April 1998 and February 2003, 
$3.45 \times 10^8$ events with $N_{\mbox{\small hit}} \ge 4$ have been recorded. 
For this analysis we used 22597 events with hit channel multiplicity
$N_{\mbox{\small hit}}>$15 which  obey the condition: 
\begin{equation}
\label{eq1}
t_{\mbox{\footnotesize min}}=\mbox{\footnotesize min}(t_i-t_j)>-10 \,\, \mbox{ns}, \,\,\, i<j.
\end{equation}
Here, $t_i, \, t_j$ are the arrival times at channels $i,j$ and
the numbering of channels rises from top to bottom along the string.

\begin{figure*}
\includegraphics*[width=.5\textwidth,height=6.0cm]{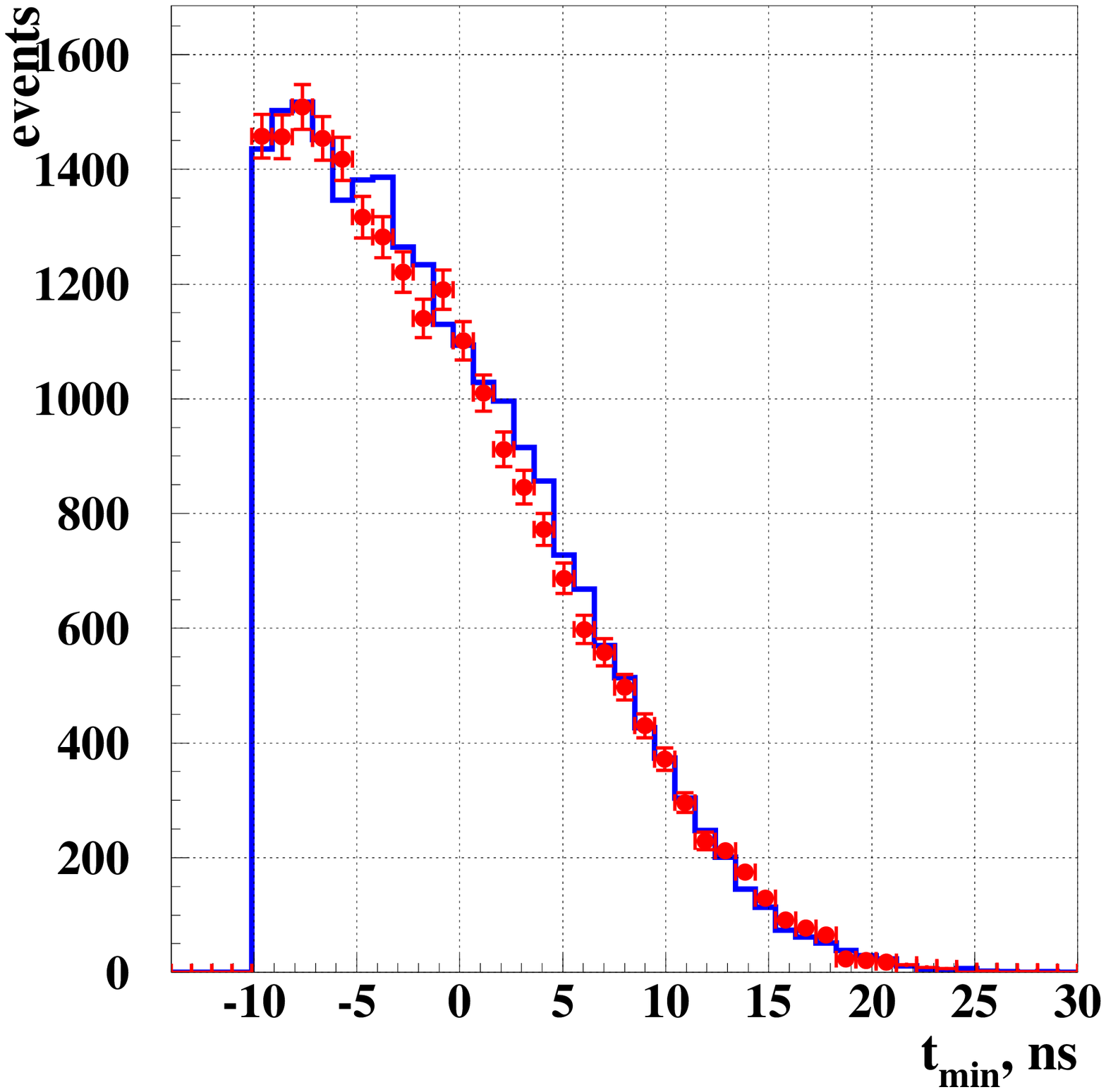}
\hfill
\includegraphics*[width=.5\textwidth,height=5.4cm]{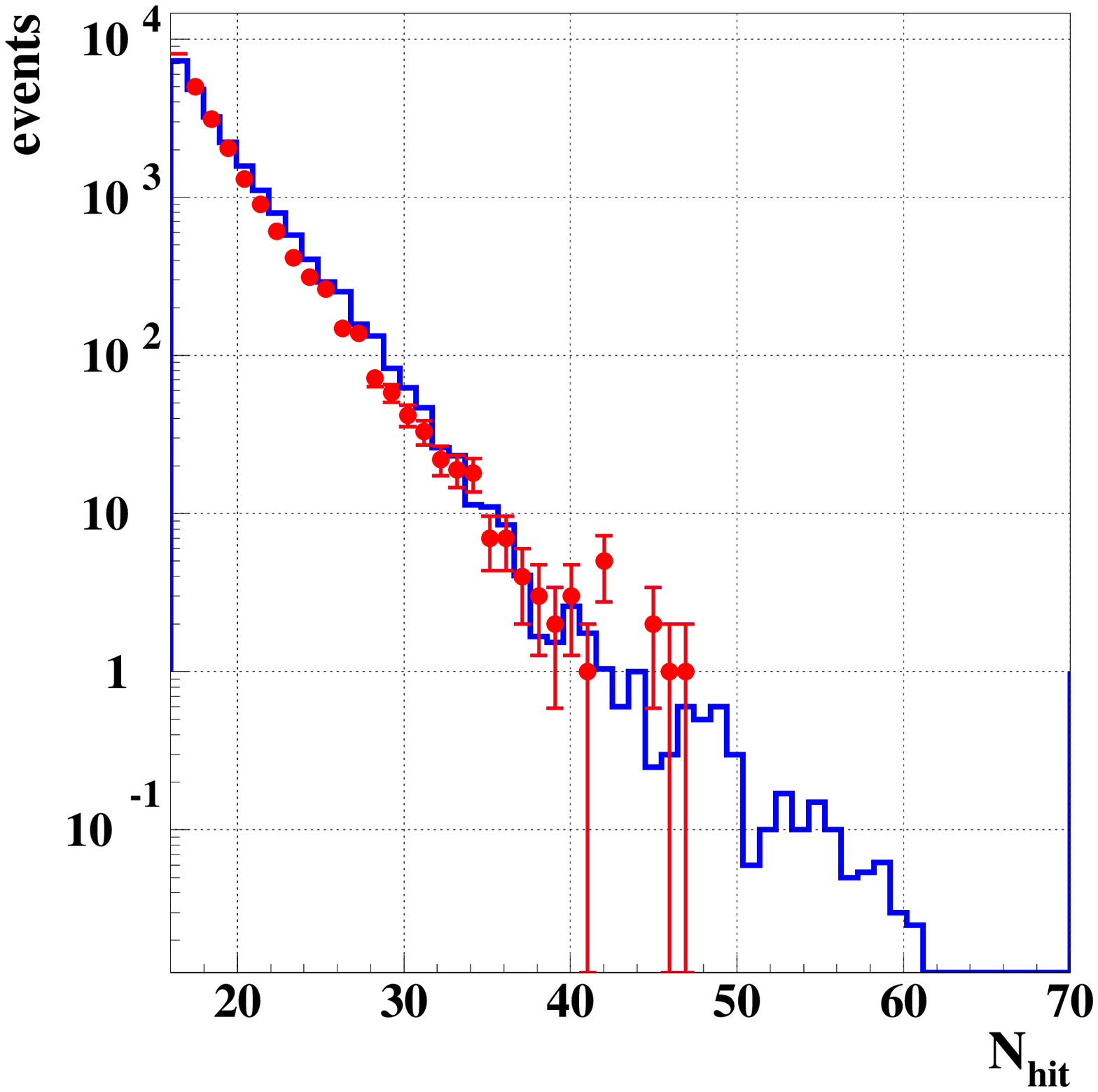}
\\
\vspace*{-3mm}
\caption{Left panel: 
the $t_{\mbox{\footnotesize min}}$ distribution of experimental events (dots)
which survive condition (\ref{eq1}) as well as the expected 
distribution of simulated background events (histogram). Right panel:
the $N_{\mbox{\small hit}}$ distribution of experimental events (dots)
as well as background prediction (histogram).}
\label{fig11}
\end{figure*}

Figure \ref{fig11} shows the $t_{\mbox{\footnotesize min}}$ and $N_{\mbox{\small hit}}$ 
distributions for experiment (dots) and background simulation (histograms).
The distribution of experimental events 
is consistent with the background simulation. 
No statistically significant excess above the background 
from atmospheric muons has been observed. 

With no experimental events outside the area populated by background 
events in the ($t_{\mbox{\footnotesize min}}, 
N_{\mbox{\small hit}}$)-parameter space,
we derive upper limits on the fluxes of high energy neutrinos as
predicted by different models of neutrino sources. 

The detection volume $V_{\mbox{\footnotesize eff}}$ averaged over all neutrino arrival directions,
rises from $\sim$10$^5$ m$^3$ for 10 TeV up to 
(4-6)$\times$10$^6$ m$^3$ for $10^4$ TeV and significantly exceeds the 
geometrical volume $V_{\mbox{\footnotesize g}}\approx$ 10$^5$ m$^3$ of {\it NT200}. This is
due to the low light scattering and the nearly not dispersed light fronts
from Cherenkov waves originating far outside the geometrical
volume. 

Since no event has been observed which fulfills the selection conditions,
upper limits on the diffuse flux of extraterrestrial neutrinos are
calculated. For a 90\% confidence level an upper limit, $n_{90\%}=$2.5, on 
the number of signal events is obtained according to Conrad et al. 
\cite{CONRAD} with the unified Feldman-Cousins ordering \cite{FC}. We assume 
an uncertainty in signal detection of 24\% and a background of zero events 
(which leads to a conservative estimation of $n_{90\%}$ according to 
the Feldman-Cousins approach).
If the expected numbers of signal events $N_{\mbox{\footnotesize model}}$ 
is larger than $n_{90\%}$, the model is ruled out at 90\% CL.
Table \ref{tab5} represents event rates and model rejection factors (MRF) 
$n_{90\%}/N_{\mbox{\footnotesize model}}$ 
for models of astrophysical neutrino sources 
obtained from our search. Recently, similar results have been 
presented by the AMANDA collaboration \cite{AMANDAHE,AMANDAMU}. 
Model rejection factors obtained by AMANDA are also shown in 
Table \ref{tab5}.
\begin{table*}[htb]
\caption{Event rates and model rejection factors for models of
astrophysical neutrino sources. The assumed upper limit on the
number of signal events with all uncertainties incorporated is $n_{90\%}=2.5$}
\label{tab5}
  \begin{tabular}{@{}lccccc|c}
\hline
 & \multicolumn{5}{c|}{BAIKAL } & AMANDA \cite{AMANDAHE,AMANDAMU}\\
\hline
Model & $\nu_e$ & $\nu_{\mu}$ & $\nu_{\tau}$ & $\nu_e+\nu_{\mu}+\nu_{\tau}$ & $n_{90\%}/N_{\mbox{\footnotesize model}}$ & $n_{90\%}/N_{\mbox{\footnotesize model}}$  \\
\hline
  10$^{-6}\times E^{-2}$ & 1.33 & 0.63 & 1.12 & 3.08 & 0.81 & 0.86  \\
  SS Quasar \cite{SS} & 4.16 & 2.13 & 3.71 & 10.00 & 0.25 & 0.21  \\
  SP u  \cite{SP}& 17.93 & 7.82 & 14.43 & 40.18 & 0.062 & 0.054  \\
  SP l \cite{SP}& 3.14 & 1.24 & 2.37 & 6.75 & 0.37 & 0.28  \\
  P $p\gamma$ \cite{P}& 0.81 & 0.53 & 0.85 & 2.19 & 1.14 & 1.99  \\
  M $pp+p\gamma$ \cite{M} & 0.29 & 0.22 & 0.35 & 0.86 & 2.86 & 1.19  \\
  MPR \cite{MPR}& 0.25 & 0.14 & 0.24 & 0.63 & 4.0 & 4.41  \\
  SeSi \cite{SeSi} & 0.47 & 0.26 & 0.44 & 1.18 & 2.12 & -  \\
  \hline
\end{tabular} 
\end{table*}
The models by Stecker and Salamon \cite{SS} labeled ``SS Q'',
as well as the models by Szabo and Protheroe \cite{SP} ``SP u''
and ``SP l'' represent models for neutrino production in the central
region of Active Galactic Nuclei 
and are ruled out with $n_{90\%}/N_{\mbox{\footnotesize model}} 
\approx$ 0.06 - 0.4.
Further shown are models for neutrino production in AGN jets:
calculations by Protheroe \cite{P}, by Mannheim \cite{M},
by Mannheim, Protheroe and Rachen
\cite{MPR} (model ``MPR'') and 
by Semikoz and Sigl \cite{SeSi} ``SeSi''.
The latter models for blazars are currently not excluded.
For an $E^{-2}$ behaviour of the neutrino spectrum and a flavor ratio 
$\nu_e:\nu_{\mu}:\nu_{\tau}=1:1:1$, the 90\% C.L. upper limit on the 
neutrino flux of all flavors obtained with the Baikal neutrino telescope  
{\it NT200} (1038 days) is:
\begin{equation}
E^2\Phi<8.1 \times 10^{-7} 
\mbox{cm}^{-2}\mbox{s}^{-1}\mbox{sr}^{-1}\mbox{GeV}.
\label{eq2}
\end{equation}
Assuming an upper limit on the number of signal events $n_{90\%}=$2.5,
the model-independent limit on $\bar{\nu_e}$ at the W - resonance energy is: 
\begin{equation}
\Phi_{\bar{\nu_e}} < 3.3 \times 10^{-20}
\mbox{cm}^{-2}\mbox{s}^{-1}\mbox{sr}^{-1}\mbox{GeV}^{-1}.
\label{eq3}
\end{equation}
\begin{figure*}
\includegraphics*[width=.45\textwidth,height=6.0cm]{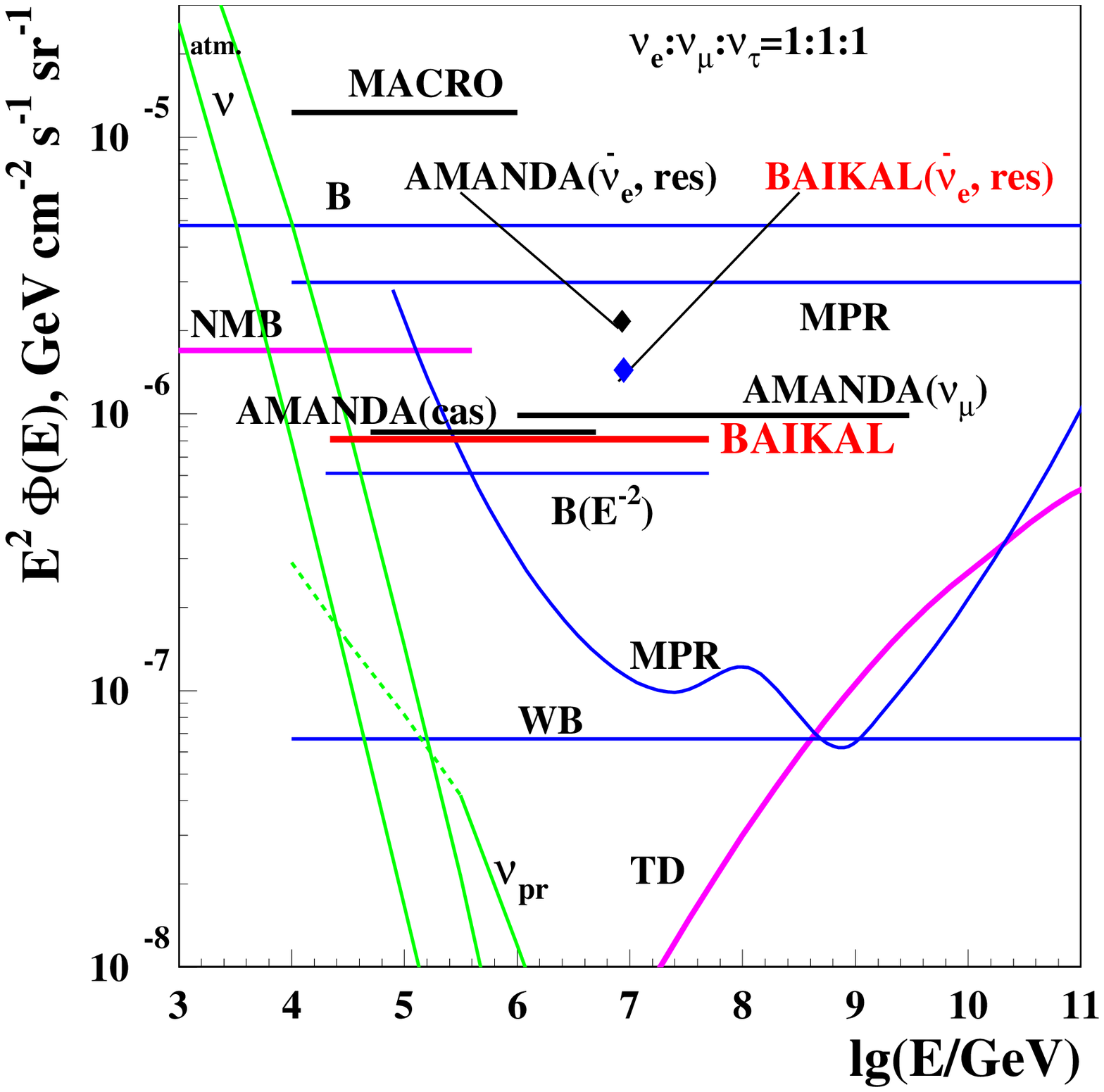}
\hfill
\includegraphics*[width=.45\textwidth,height=6.0cm]{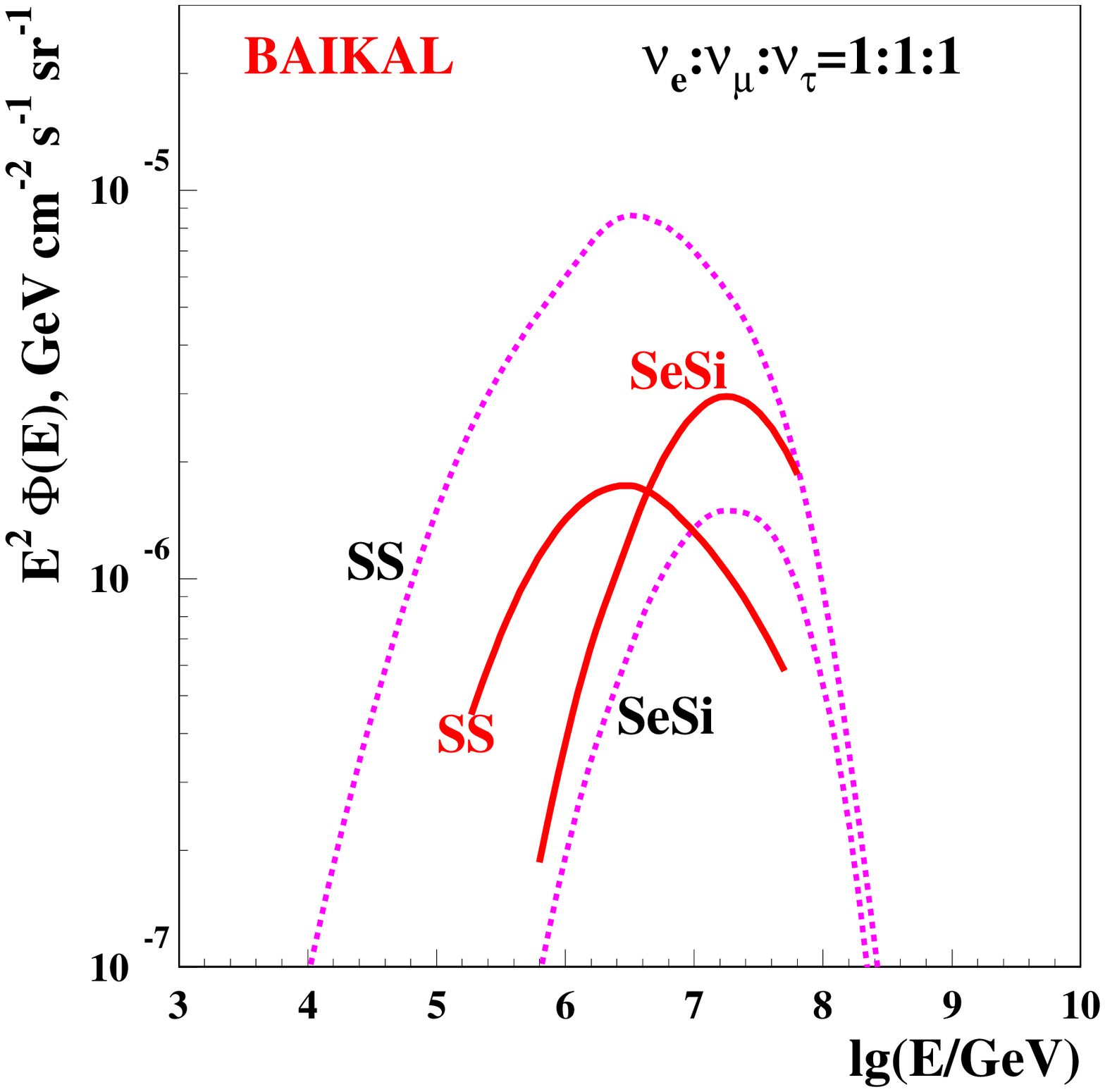}
\\
\vspace*{-5mm}
\caption{Left panel: neutrino flux 
predictions in different models of neutrino sources 
compared to experimental upper limits to $E^{-2}$ fluxes obtained
by various experiments.
(see text).
Right panel: 
experimental limits compared to two
model predictions. Dotted curves: predictions
from model SS \cite{SS} and SeSi \cite{SeSi}. Full curves:
upper limits to spectra of the same shape.
Model SS is excluded (MRF=0.21), model SeSi is not
(MRF=2.12).
}
\label{fig16}
\end{figure*}
Figure \ref{fig16} (left panel) shows our upper limit on the
($\nu_e + \nu_{\mu} +\nu_{\tau}$) $E^{-2}$ diffuse flux
as well as the model independent limit on the resonant $\bar{\nu}_e$ flux 
(diamond). Also shown are the limits obtained by AMANDA and MACRO 
\cite{AMANDAHE,AMANDAMU,MACROHE}, theoretical bounds obtained by 
Berezinsky (model independent (B) and for an $E^{-2}$ shape
of the neutrino spectrum (B($E^{-2}$)) 
\cite{Ber3}, by Waxman and Bahcall (WB) \cite{WB1}, by Mannheim et al.(MPR) 
\cite{MPR}, predictions for neutrino fluxes from topological defects (TD) 
\cite{SeSi}, prediction on diffuse flux from AGNs according to Nellen et al. 
(NMB) \cite{NMB}, 
as well as the atmospheric 
conventional neutrino \mbox{fluxes \cite{VOL}} from horizontal and vertical 
directions (upper and lower curves, respectively) and atmospheric prompt 
neutrino fluxes obtained by Volkova et al. \cite{VPPROMPT}.
Our upper limits (solid curves) on diffuse fluxes
from AGNs shaped according to the model of Stecker and 
Salamon (SS) \cite{SS} and of Semikoz and Sigl (SeSi) \cite{SeSi} 
are shown in the right panel of fig. \ref{fig16}. 

In March/April 2005 we fenced a large part of the
search volume 
with three sparsely
instrumented strings
(see \cite{IC05_stat} for details).
The three-year sensitivity of this
enlarged detector {\it NT200}$+$, with about 5 Mton enclosed
volume, is approximately 
$E^2 \Phi_{\nu_e} \sim 10^{-7}$cm$^{-2}$s$^{-1}$sr$^{-1}$GeV
for $E>$10$^2$ TeV, i.e. three-four times better than {\it NT200}.
{\it NT200}$+$ will search for neutrinos from AGNs, GRBs
and other extraterrestrial sources, neutrinos from cosmic ray
interactions in the Galaxy as well as high energy atmospheric muons
with $E_{\mu}>10$ TeV.

{\it This work was supported by the Russian Ministry of Education and Science,
the German Ministry of Education and Research and the Russian Fund of Basic 
Research} ({\it grants} \mbox{\sf 05-02-17476}, \mbox{\sf 04-02-17289} 
{\it and} \mbox{\sf 02-07-90293}), {\it and by the Grant of President of 
Russia} \mbox{\sf NSh-1828.2003.2}. 

\end{document}